\documentclass[11pt]{article} 

\def\be{\begin{equation}}\def\ee{\end{equation}}

\let\a=\alpha \let\b=\beta  \let\g=\gamma  \let\d=\delta \let\e=\varepsilon
\let\z=\zeta       
\let\m=\mu        \let\x=\xi         
\let\s=\sigma \let\t=\tau   \let\f=\varphi

\font\tenmib=cmmib10%
\font\sevenmib=cmmib7%
\font\fivemib=cmmib5%
\mathchardef\Bx   = "0518  %xi bold greek
\def\ie{{\it i.e.\ }}

\title{\bf Thermostats, chaos and Onsager reciprocity}
\author{\small\textsc{Giovanni Gallavotti}%
\\
\small Dipartimento di Fisica and INFN\\
\small Universit\`a di Roma 
{\it La Sapienza}\\
\small P.~A.~Moro 2, 00185, Roma, Italy\\
\small \texttt{giovanni.gallavotti@roma1.infn.it}
}
\date{\today}
\newdimen\xshift \newdimen\xwidth \newdimen\yshift \newdimen\ywidth

\def\ins#1#2#3{\vbox to0pt{\kern-#2pt\hbox{\kern#1pt #3}\vss}\nointerlineskip}

\def\eqfig#1#2#3#4#5{
\par\xwidth=#1pt \xshift=\hsize \advance\xshift
by-\xwidth \divide\xshift by 2
\yshift=#2pt \divide\yshift by 2
{\hglue\xshift \vbox to #2pt{\vfil
#3 \includegraphics{#4.eps}
}\hfill\raise\yshift\hbox{#5}}}
\begin{document}
\maketitle
\begin{abstract}{\it Finite thermostats are studied in the context of
    nonequilibrium statistical mechanics. Entropy production rate has
    been identified with the mechanical quantity expressed by the
    phase space contraction rate and the currents have been linked to
    its derivatives with respect to the parameters measuring the
    forcing intensities. In some instances Green-Kubo formulae, hence
    Onsager reciprocity, have been related to the fluctuation
    theorem. However, mainly when dissipation takes place at the
    boundary (as in gases or liquids in contact with thermostats),
    phase space contraction may be independent on some of the forcing
    parameters or, even in absence of forcing, phase space contraction
    may not vanish: then the relation with the fluctuation theorem
    does not seem to apply. On the other hand phase space contraction
    can be altered by changing the metric on phase space: here this
    ambiguity is discussed and employed to show that the relation
    between the fluctuation theorem and Green-Kubo formulae can be
    extended and is, by far, more general.}\end{abstract} \*
\vskip3mm

I am honored to be given this occasion to thank J\"urgen Fr\"olich and
Tom Spencer for the time they spent discussing with me and
communicating their insights, projects and ideas. In particular
I have drawn inspiration from the one among the two to whom I have
been in closer contact, (JF), particularly at the time when he worked at
IHES: his strong criticism of any ``mathematical deviations'' from
physical problems has been always extremely effective on my work.

\section{Thermostats}\label{sec1}\setcounter{equation}{0}

A mechanical interpretation of the entropy production rate in
nonequilibrium systems interacting with thermostats and possibly
subject to external non conservative (``stirring'') forces has emerged
from simulations and studies on nonequilibrium statistical mechanics
since the early 1980's, \cite{No84,Ho85,EM90,Ga08}. It is interpreted
as phase space contraction rate, as defined by the divergence of the
equations of motion which we write symbolically $\dot x=f(x)$, \ie
$\s(x)=-\sum_i \partial_{x_i} f(x)$.

General thermostats acting on a mechanical system, on which also
external non conservative forces may act, will be modeled as described
in Fig.1 and illustrated in the caption, \cite{Ga08,Ga07b}:

\eqfig{302}{85}{\ins{90}{60}{${\bf X}_0,{\bf X}_1,\ldots,{\bf X}_n$}
\ins{60}{27}{$\ddot{{\bf X}}_{0i}=-\partial_i U_0({\bf X}_0)-\sum_{j}
\partial_i U_j({\bf X}_0,{\bf X}_j)+{\bf E}_i({\bf X}_0)$}
\ins{60}{10}{$\ddot{{\bf X}}_{ji}=-\partial_i U_j({\bf X}_j)-
\partial_i U_j({\bf X}_0,{\bf X}_j)-\a_j {\bf{\dot X}}_{ji}$}
}{fig1}{(fig1)\label{fig1}}
\*

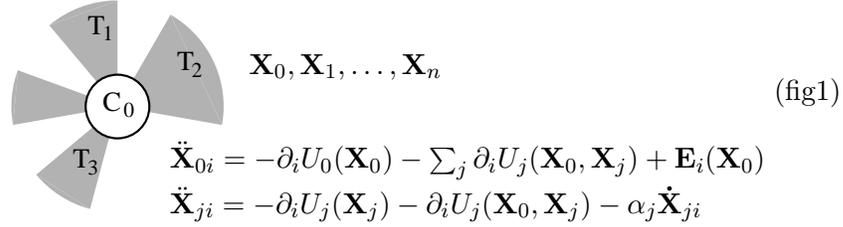
\noindent{}{Fig.1: \small\ The $1+n$ boxes $C_0,T_1,\ldots,T_n$ contain
  $N_0,N_1,\ldots, N_n$ particles, of mass $m=1$, whose positions and
  velocities are denoted ${\bf X}_0,{\bf X}_1,\ldots,{\bf X}_n$, and
  $\dot{\bf X}_0,\dot{{\bf X}}_1,\ldots,\dot{{\bf X}}_n$
  respectively. The ${\bf E}$ denote external, non conservative,
  forces and the multipliers $\a_j$ model the thermostats and are so
  defined that the kinetic energies $K_j=\frac12 \dot{{\bf X}}_j^2$
  are exactly constants of motion with values $K_j=\frac32 N_j k_B
  T_j$, $k_B=$ Boltzmann's constant, $j=1,\ldots,n$.  The energies
  $U_0,U_j,W_{0,j},\,j>0,$ should be imagined as generated by pair
  potentials $\f_0,\f_j,\f_{0,j}$ short ranged, smooth, or with a
  singularity like a hard core.\footnote{\small Singularities of
  different type but care has to be excercised in formulating and by
  external potentials modeling the containers walls and for simplicity
  the assumption of smoothness (possibly in presence of a hard core)
  is made here. For the more general cases, like Lennard-Jones
  potentials, see \cite{BGGZ05}.}}

\* To imply $\dot K_j=0$ in the above model the multiplier $\a_j$ has
to be $\a_j=-\frac{(Q_j+\dot U_j)}{3 N_j k_B T_j}$, where

\be Q_j{\buildrel def\over=}-\dot{\bf X}_j\cdot\partial_{{{\bf X}}_j}
W_{0,i}({\bf X}_{0},{{\bf X}}_j)\label{e1.2}\ee
is naturally interpreted as the {\it heat} ceded per unit time to the
thermostat $C_j$. The phase space contraction rate, neglecting for
simplicity $O(N_j^{-1})$, is computed from the equation in Fig.1 to be

\be\s(X)=\sum_j\frac{Q_j+\dot U_j}{k_B T_j}\label{e1.1}\ee
(each addend should be multiplied by the factor $(1-\frac2{3N_j})$ if
$O(N_j^{-1})$ is not neglected).

Of course $\s(x)$ depends upon the metric used on phase space and on
the density giving the volume element: both are arbitrary and
Eq.(\ref{e1.1}) yields the contraction rate for the Euclidean metric
and density $1$: \ie for the Liouville volume. Because of such
ambiguity $\s(x)$ {\it cannot} have an immediate physical
meaning. However its time average, and the fluctuations of its finite
time averages over long time intervals, have an intrinsic meaning,
independent of the choices of the metric and the density,
\cite{Ga07b}, at least if the motions are ``chaotic'', see below.

Some interesting concrete examples of the above systems are
illustrated in Fig.2 and Fig.3.

\eqfig{290}{70}{
\ins{24}{63}{${{\bf E}}\ \to$}
\ins{70}{50}{periodic boundary (``{\it wire}'')}
\ins{70}{32}{$m\ddot{\bf x}={\bf E} -\a \dot{\bf x}$}
}{fig2}{(fig.2)}
\*

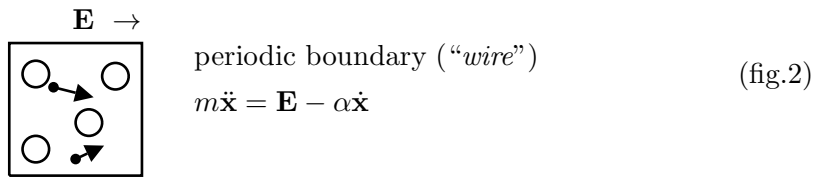
\noindent{}{Fig2. \small A modern version of the classical Drude's model for
  electric conductivity.}

\*
\noindent{}In Fig.2 $\a=\frac{{\bf E}\cdot\dot{\bf x}}{m\dot{\bf x}^2}$ and this is
an electric conduction model of $N$ charged particles ($N=2$ in the
figure) in a constant electric field ${\bf E}$ and interacting with a
lattice of obstacles (circles in the figure); it is
``autotermostatted'' (because $C_0$ and $T_1$ coincide) in $2$
dimensions. This is a model that appeared since the early days (Drude,
1899, \cite{Be64}) in a slightly different form (\ie in dimension $3$
and with the thermostatting realized by replacing the $-\a\dot{\bf x}$
force with the prescription that after collision with an obstacle
velocity is rescaled to $|\dot{\bf x}|=\sqrt{\frac{3}m k_B T}$. 

The thermostat forces are a model of the effect of the interactions
between the particle (electron) and a background lattice (phonons).
This model is remarkable because it is the first nonequilibrium
problem that has been treated with real mathematical attention and for
which the analog of Ohm's law for electric conduction has been proved
if $N=1$, \cite{CELS93}.

Another example is a model of thermal conduction, Fig.3:

\eqfig{360}{60}{
\ins{90}{60}{$T_1$}
\ins{170}{61}{$C_0$}
\ins{250}{61}{$T_2$}
}{fig3}{(fig.3)\label{fig3}}
\*

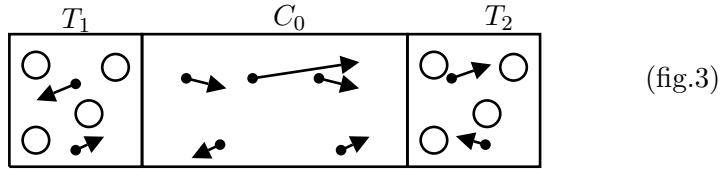
\noindent{}{Fig3. \small A model for thermal and electric conduction.}
\*

\noindent{}in which $N_0$ hard disks interact by elastic collisions with each
other and with other hard disks ($N_1=N_2$ in number) in the
containers labeled by their temperatures $T_1,T_2$: the latter are
subject to elastic collisions between themselves and with the disks in
the central container $C_0$; the separation reflect elastically the
particles when their {\it centers} reach them, thus allowing interactions
between the thermostats and the main container particles.
Interactions with the thermostats take place only near the separating
walls.  

If one imagines that the upper and lower walls of the central
container are identified (realizing a periodic boundary condition) and
that a constant field of intensity $E$ acts in the vertical direction
then two forces conspire to keep it out of equilibrium, and the
parameters ${\bf F}=(T_2-T_1,E)$ characterize their strength: matter and
heat currents flow.

The case $T_1=T_2, E\ne0$ has been studied in simulations to check that the
thermostats are ``efficient'': \ie that the simple interaction, via
collisions taking place across the boundary, is sufficient to allow the
systems to reach a stationary state, \cite{GG07}.

Thermostat models similar to the above have been considered in the
literature, \cite{EM90,ECM93,GC95}. A fundamental problem with the
model in Fig.1 is that it is not clear which detailed assumptions have
to be made on the interactions to insure that almost all initial
conditions evolve staying in a bounded region in phase space so that
they can be expected to determine a stationary state. This can be
called the ``thermostat efficiency problem'' and it is, for
nonequilibrium, the analogue of the Hamiltonian stability problem in
equilibrium, \cite{Ga00}.  The experiment in \cite{GG07} encourages
the idea that the assumptions could be very general and fairly simple.
In \cite{Ga96} a model like the one in Fig.3 was studied but the
confinement difficulty was avoided by requiring that also the total
kinetic energy $K_0$ in the central container was constant thanks to
an extra thermostatting force $-\a_0 \dot{{\bf X}}_0$ with a properly
chosen $\a_0$.

The model in Fig.3 {\it without} thermostatting forces to
keep $K_j,\,j>0$ constant, hence with a purely Hamiltonian evolution,
has been carefully studied in \cite{Ja99} which also gives the
clearest account on the so called ``transient fluctuation theorem''
improving and extending its earlier formulation in \cite{BK81a}, and
obtains implicitly also a transient version of the result on
fluctuation patterns, analogous to the one derived earlier for steady
states in \cite{Ga99}.

In \cite{Ja99} there is also a careful analysis of the model in Fig.3
with the aim of obtaining results for stationary states:
stationarity is made possible by taking the thermostats infinitely
large stressing the (formidable) problems that one should encounter in
attempting a rigorous proof.

In this paper (and in all my preceding ones) I have chosen to consider
only finite thermostats with empirical thermostat forces and studied a
few problems by introducing a single assumption, the chaotic
hypothesis.

\section{Chaos}\label{sec2}\setcounter{equation}{0}

Microscopic motions are in all possible empirical senses
``chaotic''.  The paradigm of chaotic motions are the hyperbolic
transitive systems: these are smooth systems whose evolution can be
intuitively described by saying that each phase space point moves
being seen by the comoving neighboring points as a hyperbolic fixed point.

Another intuitive way to look at such systems is to say that the phase
space points can be coded into sequences $\Bx=(\x_i)_{i=-\infty}^{\infty}$
of symbols, say the digits $0,1,2,$ $\ldots,q<\infty$, in such a way that
the dynamics becomes the trivial shift of the sequence $\Bx$, and all
sequences which satisfy $M_{\x_i,\x_{i+1}}\equiv1$ represent one phase
space point, $M$ being a ``compatibility matrix'' with elements
$M_{ij}=0,1$ which is transitive (\ie $M^s_{ij}>0$ for some
$s$). There may be ambiguities, \ie different sequences may represent
the same point, but this can happen on a zero volume set of points
only, in close analogy with the familiar ambiguity in the
representation of number by digits (where $0.9999..$ and $1.0000...$
are the same number).

It is natural, at least for some \cite{Ru78b,GC95,Ru98}, to imagine
that motions of complex systems, like gases or liquids, are chaotic in
the simplest sense (which is also the strongest) of being hyperbolic
transitive on the attracting sets (also called Anosov systems). The
{\it chaotic hypothesis}, proposed in \cite{GC95}, see also
\cite{Ga00}, reflects this remark.

\*

\noindent{}{\bf Chaotic hypothesis} {\it Attracting sets for mechanical systems
are smooth surfaces on which motion is smooth, hyperbolic and transitive.}  \*

\noindent{}This is an hypothesis that has to be considered in the same sense
as the ergodic hypothesis for equilibrium statistical mechanics,
\cite{Ru73}.  Hence {\it it might be at first disturbing}. 

However disturbing assumptions are common in the literature and,
nevertheless, are often fruitful. I just mention the assumption of
periodicity with equal period (``monocyclicity'') of the motions of
mechanical systems: it was employed in the derivation of the second
law from the action principle in Boltzmann, \cite{Bo866}: this
assumption was considered also by Clausius, Maxwell, Helmholtz and was
the basis of the early works on the mechanical interpretation of the
second law, \cite{Cl871,He884a}. At the time there must have been
objections to such a bold assumption and someone must have declared,
as it was done a little later about its modification into the ergodic
hypothesis (and as it is done today about the chaotic hypothesis),
that it is ``a strong assumption as the periodicity (or ergodicity)
hypothesis raises the question of which systems of practical interest
are ``periodic'' (``ergodic''), since almost none of them is actually
such'', see \cite{MR007}. Similar statements can be found in the
literature, even in good papers.

Chaotic systems (in the above sense) admit a statistics (called
SRB statistics, \cite{Si68a,Bo70a,BR75}), \ie a probability
distribution $\m$ on each attracting set which, by integration, gives
the average values of the observables $G(x)$ on trajectories whose
initial data $x$ are randomly chosen, near enough to an
attracting set, with a distribution with some (arbitrary) density:

\be \langle{G}\rangle\,=
\lim_{T\to\infty}\frac1T\sum_{j=0}^{T-1} G(S^jx)=\int
G(y)\m(dy),\qquad \hbox{with probability $1$}\label{e2.1}\ee
where $x\to Sx$ is a discretized time evolution map, obtained by
timing observations on the occurrence of some selected event. Or in
the (unphysical, yet customary and interesting) case of observations
in continuous time

\be \langle{G}\rangle\,=\lim_{T\to\infty}\frac1T \int_0^T G(S_t x)\,dt =\int
G(y)\m(dy),\qquad \hbox{with probability $1$}\label{e2.2}\ee
where $x\to S_tx$ denotes the evolution of the initial data $x$ via
the equations of motion, \cite{Ge98}.

If motion is chaotic (\ie hyperbolic, regular, transitive) the finite
time averages

\be \g=\langle{G}\rangle_\t= \frac1\t\sum_{j=0}^{\t-1} G(S^jx)\label{e2.3}\ee
satisfy a {\it large deviations law}, \ie fluctuations off the average
$\langle{G}\rangle$ as large as $\t$ itself are controlled by a function
$\z(\g)$ convex and analytic in a (finite) interval $(\g_1,\g_2)$, 
maximal at $\langle{G}\rangle$. This means that the probability that $\g\in
[a,b]$ satisfies

\be P_\t(\g\in [a,b])\simeq \, e^{\t\,\max_{[a,b]}\z(\g)},\qquad \forall
a,b\in (\g_1,\g_2)\label{e2.4}\ee
and the interval $(\g_1,\g_2)$ is non trivial if
$\langle{G^2}\rangle-\langle{G}\rangle^2>0$, 
\cite{Si68a,Si77,GBG04}.  If $\z(\g)$ is
quadratic at its maximum (\ie at $\langle{G}\rangle$) then this implies a
central limit theorem for the fluctuations of
$\sqrt{\t}\,\langle{G}\rangle_\t$, but Eq.(\ref{e3.4}) is a much stronger
property.

\* 
\noindent{}{\it Remarks:} (1) The hypothesis holds also in equilibrium; if
the system admits a dense trajectory in phase space it implies the
classical ergodic hypothesis.
\\
(2) If the observable $G$ has nonzero SRB-average
it is convenient to consider instead the observable
$\frac{G}{\langle{G}}\rangle$ because it is dimensionless, just as in the case
of $\langle{G}\rangle=0$ it is convenient to consider the dimensionless
observable $\frac{G}{\sqrt{\langle{G^2}\rangle}}$.
\\ (3) If the dynamics is {\it reversible}, \ie there is a smooth,
isometric, map $I$ of phase space such that $I^2=1$ and $IS_t=S_{-t}
I$ or in the discrete case $IS=S^{-1}I$, then any {\it time reversal odd}
observable $G$, with non zero average and nonzero dispersion
$\langle{G^2}\rangle-\langle{G}\rangle^2>0$, is such that the interval of
$(\g_1,\g_2)$ of large deviations for $\frac{G}{\langle{G}\rangle}$ is at
least $(-1,1)$ provided there is a dense orbit (which also implies
existence of only one attracting set).
\\ (4) The systems in the thermostats model of Sec.\ref{sec1} are all
reversible with $I$ being the ordinary time reversal, change in sign of
velocity with positions unaltered, and the phase space contraction
$\s(x)$ is odd under time reversal, see Eq.(\ref{e1.1}). Therefore if
$\s_+=\langle{\s}\rangle>0$ it follows that the observable

\be p'=\frac1\t\sum_{j=0}^{\t-1}
\frac{\s(S^jx)}{\s_+}\label{e2.5}\ee
has domain of large deviations of the form $(-\overline g,\overline g)$
and contains $(-1,1)$.  
\\ 
(5) Since by Eq.(\ref{e1.1}) $\s$ differs from
$\e(x)=\sum_{j>0}\frac{Q_j}{k_B T_j}$ by the time derivative of an
observable, it follows that the finite or infinite time averages $ \s$
and of $\e$ have, for large $\t$, the same distribution. Therefore the
same large deviations function $\z(p)$ controls the fluctuations of
$p'$ above and of
 
\be p=\frac1\t\sum_{j=0}^{\t-1} \frac{\e(S^jx)}{\s_+},
\qquad\s_+\equiv\langle{\s}\rangle_{SRB}=\langle{\e}\rangle_{SRB},\label{e2.6}\ee
and it has been shown, \cite{GC95,GC95b} and in a mathematical form in
\cite{Ga95b}, that under the chaotic hypothesis and reversibility of
motions on the attracting set, the function $\z(p)$ has the
symmetry property

\be \z(-p)=\z(p)-p\s_+, \qquad\hbox{\rm for all}\ p\in(-\overline p,\overline
p)\label{e2.7}\ee
and $\overline p\ge1$. This is the {\it fluctuation theorem} of
\cite{GC95} (it requires a proof and therefore it should not be
confused with several identities, see for instance \cite{CG99}, with
which, for reasons that I fail to understand, it has been often
identified). The interest of the theorem is that it is {\it
universal}, model independent yielding a parameter free relation which
deals with a quantity which has the physical meaning of entropy
production rate and therefore has an independent macroscopic
definition and is accessible to experiments.
\\ (6) Eq.(\ref{e2.7}) is closely related to the theorem in
\cite{Ja99}, from which it differs only because it deals with finite
thermostats assuming the (strong) chaotic hypothesis, rather than
dealing with infinite thermostats and assuming (strong) ergodicity
properties. In spite of the latter work several paper have appeared in
the literature trying to get rid of the chaotic hypothesis without
adding much (if anything) to the lucid discussion in \cite{Ja99} about
the necessity of suitable assumptions in order to allow extending a
transient fluctuation relation (which is an identity, requiring no
assumption, on the full phase space, \cite{CG99}) to a stationary one
(which deals with properties that hold on a subset of zero probability
with respect to the initial data sampling).
\\ 
(7) The fluctuation theorem has several extensions including a
remarkable, parameter free relation that concerns the relative
probability of patterns of evolution of an observable and their
reversed patterns, \cite{Ga97,Ga00,Ga02}, related to the
Onsager--Machlup fluctuations theory, which keeps being rediscovered
in various forms and variations in the literature.

\section{Onsager reciprocity}\label{sec3}\setcounter{equation}{0}

Another consequence of the fluctuation theorem are the Onsager
reciprocity and Green-Kubo formulae for the infinitesimal deviations
from equilibrium, \cite{Ga96}; the latter can be independently derived
(in a simpler way) from the chaotic hypothesis and time reversal
symmetry assumed only at equilibrium, \cite{CELS93}, as it will be
shown in the concluding comments, or as discussed from a somewhat
different viewpoint in \cite{GR97}.

Here the aim is to show that {\it the  Green-Kubo formulae, hence Onsager's
reciprocity, can be regarded as the version at zero forcing of the
fluctuation theorem for stationary states}. 

In the case in which $T_1=T_2=\ldots=T$ and ${\bf E}={\bf0}$ the
system is in thermal equilibrium and its state is characterized by a
probability distribution $\m_0$ which invariant under the time
evolution $x\to S_tx$ generated by the equations in Fig.1. Setting
$x=({\bf X}_0,\dot{{\bf X}}_0,{\bf X}_1,\dot{{\bf X}}_1,\ldots,{\bf
X}_n,\dot{{\bf X}}_n)$ it is remarkable that the distribution can be
explicitly found, \cite{EM90}, as

$$\m_0(dx)=\, const\, e^{-\b \big(U_0({\bf X}_0)+\sum_{j>0}
 \big(U_j({\bf X}_j)+W({\bf X}_0,{\bf X}_j)\big)+K_0(\dot{{\bf X}}_0 )\big)} $$
\be\cdot\textstyle
 (\prod_{j>0}\d(K_j-\frac32 N_j T))(\prod_{j\ge0} d\dot{{\bf X}}_j\, d{\bf
 X}_j)\label{e3.1}\ee
where $\b=\frac1{k_B T}$ (neglecting $O(N_j^{-1})$ for
simplicity). Calling the ``unperturbed'' energy $H_0(x)=K_0(\dot{\bf
X}_0 )+\sum_{j\ge0} U_j({\bf X}_j) +\sum_{j>0} W_j({\bf X}_0,{\bf
X}_j)$\footnote{\small The kinetic energy of the thermostats is an
  additive constant and therefore is not explicitly written.} and
$\widetilde\d({\bf K}_j(x), T_j)= \d(K_j(\dot{{\bf X}}_j)-\frac32 N_j T_j))$, 
Eq.(\ref{e3.1}), written more compactly, is

\be\m_0(dx)= 
const\,e^{-\b H_0(x)} \,\prod_{j>0}\widetilde\d(K_j(x),T)\,dx\label{e3.2}\ee
which is a distribution in an ensemble which, for the system in $C_0$,
is equivalent to the canonical one (for $N_0,L_0\to\infty,\,
N_0/L_0^3=const$ if $L_0$ is the side of the container).

We now want to compare the average values of various currents that are
swit\-ched on when ${\bf E}$, the external forces, become non zero and the
temperatures of the thermostats become different: ${\bf E}\ne{\bf0}$ and
$T_j=T+\e_j$.  More precisely we look for the relations between
infinitesimal forcing {\it actions} and the corresponding {\it
currents}, \ie the susceptibility coefficients.

The currents are related to the average values of the derivatives of
the entropy production with respect to the forces (material currents)
or to the temperature inequalities (heat currents). {\it However} the
arbitrariness inherent in the phase space contraction generates
interesting questions: for instance in the model in Fig.1 the
phase space contraction with respect to the Liouville volume {\it is
independent of the external forces ${\bf E}$}, see Eqs.(\ref{e1.2}),(\ref{e1.1}),
so that $\partial_E\s\equiv0$, while it is {\it obvious} that the external
forces generate material currents, being non conservative.

On the other hand even in equilibrium a thermostatted system exchanges
heat with the thermostats: hence there is a production of entropy
which has a zero average but which is not zero and equal to
$\sum_j\frac{Q_j}{k_B T}$.

It is therefore interesting to see, first, why in equilibrium (\ie
when the thermostats have all the same temperature and no external
forces act) the SRB-average of $\sum_j\frac{ Q_j}{k_B T}$ vanishes,
\cite{Ru96}. This is the case because the latter quantity is the
derivative of $\b H_0(x)$. In fact the derivative $\b H_0$ is $\b$
times the work done on the system by the forces $-\a \dot{{\bf X}}_j$
which equals $\sum_{j>0} \frac{Q_j+\dot U_j}{k_B T}$. This means that
$\s(x)-\b \dot H_0(x)\equiv 0$ and therefore

\be \sum_{j>0} \frac{Q_j}{k_B T}=\b \dot H_0(x)-\sum_{j>0}
\frac{\dot U_j(x)}{k_B T}\label{e3.3}\ee
and the r.h.s. is a time derivative, hence it has $0$ time average.

When the system is out of equilibrium (\ie $T_j\not\equiv T$ and ${\bf
E}\ne {\bf0}$) the heat currents flowing into the thermostats divided by
the temperature are generated by the entropy production rate
$j_k(x)=\partial_{T_k} \s(x)$, while the material currents through the
system are defined by minus the derivatives with respect to the acting
forces of the work per unit time that they do, given by the
corresponding derivatives of $\dot H_0$. Thus given arbitrarily $\b$
the quantity

\be \overline \s(x)=\s(x)-\b \dot H_0(x)\label{e3.4}\ee
generates all currents up to a proportionality factor (here $\b$ is
arbitrary). It can be computed as

\be \overline\s(x)=\sum_{j>0} \frac{Q_j+\dot U_j}{k_B T_j}-\b\, {\bf
E}\cdot \dot {{\bf X}}_0-\b\,\sum_{j>0}(Q_j+\dot U_j)\label{e3.5}\ee
because, by the equations in Fig.1, $\dot H_0={\bf E}\cdot
\dot{{\bf X}}_0-\sum_{j>0}\a_j \dot{{\bf X}}_j^2$ and therefore
$\dot H_0={\bf E}\cdot
\dot{{\bf X}}_0+\sum_{j>0}(Q_j+\dot U_j)$, see Eq.(\ref{e1.2}).

Hence, discarding the time derivatives terms involving the $\dot U_j$
(parameters independent), the currents (at infinitesimal forcing) can
be generated by the function

\be\s_0(x)=\sum_{j>0} Q_j(\frac1{k_B T_j}-\frac1{k_B
  T})-{\bf E}\cdot\dot{{\bf X}}_0\frac1{k_B T}\label{e3.6}\ee 

The generating function $\s_0$ is odd under time reversal and {\it
vanishes at equilibrium} $T_j=T_i,{\bf E}={\bf0}$ if $T$ is chosen $T=T_j$;
its derivatives with respect to the forcing parameters $T_j,E_k$
generate the heat and material currents and, at the same, time
$\s_0(x)$ differs from the phase space contraction by a time
derivative.

Note that $\overline \s$ is also the phase space contraction of the volume in
phase space, {\it provided} the latter is measured by the distribution 

\be\overline\m(dx)= 
const\,e^{-\b H_0(x)} \,\prod_{j>0}\widetilde\d(K_j(x),T_j)\,dx\label{e3.7}\ee

In \cite{Ga96a} a reversible system (like the model in Fig.1), has
been considered in which the generating function for the currents {\it
$\s_0$ vanishes for vanishing ``thermodynamic forces'' ${\bf F}=
(T_1-T,\ldots,T_n-T, E_1,\ldots ,E_q)={\bf0}$} and satisfies the
fluctuation relation or, better, its extension in
\cite[Eq.(14)]{Ga96a}, has been considered. 

And it has been shown, \cite{Ga96a}, that the products of the currents,
generated by the thermodynamic forces, times $\b=\frac1{k_B T}$ , and
defined by

\be j_m=\partial_{F_r}\overline\s(x)\equiv \partial_{F_r}\s_0(x)\label{e3.8}\ee
are such that their averages $J_m=\langle{j_m}\rangle_{SRB}$ have
susceptibilities $L_{mp}=\partial_{F_m} J_p\Big|_{{\bf F}={\bf0}}$ which
satisfy

\be L_{mp}=
\frac12\int_{-\infty}^\infty 
dt\, \big(\langle j_m(S_t\cdot) j_p(\cdot)\rangle_{SRB}-
\langle j_m\rangle_{SRB}\langle j_p\rangle_{SRB}
\big)\big|_{{\bf F}={\bf0}}\label{e3.9}\ee

If the parameter $\b$ is properly chosen as mentioned above, \ie
$\b=\frac1{k_BT}$ (and only if so chosen), $\s_0$ will vanish when ${\bf
F}={\bf0}$.  Since $\overline\s$ and $\s_0$ differ by a time derivative they
can be interchangeably used in the theory of the SRB distribution and
therefore $\s_0$ satisfies the fluctuation theorem (because $\overline\s$
does); the assumptions in the derivation in \cite{Ga96a} apply and
therefore Eq.(\ref{e3.9}) yields Onsager's reciprocity
$L_{mp}=L_{pm}$, and Green--Kubo formula.

\section{Work and entropy theorems. Comments}\label{sec4}
\setcounter{equation}{0}

(1) This extends considerably the results in \cite{Ga96a,Ga96}
removing the restriction on the phase space contraction to be the
generating function of the currents.  The key is that the phase space
contraction is only defined up to a time derivative of an observable
and the generating function of the currents coincides with the phase
space contraction only if the observable is properly chosen.  \*

\noindent{}(2) it is worth stressing that the extension of the fluctuation
    theorem needed to derive from it Onsager reciprocity is an
    important one: in \cite{Ga97} it was further extended to show
    ({\it conditional reversibility theorem}) that there is a simple
    relation between the probability that an observable $F(x)$, even
    or odd under time reversal (for simplicity), follows in a time
    interval ${-\t,\t}$ a ``pattern'' $F(S_t x)=\f(t)$ or the
    ``reversed pattern'' $F(S_t x)=\f(-t)$ {\it provided} the entropy
    production rate is fixed, \cite{Ga97}. A statement that can be
    colorfully quoted as {\it .. relative probabilities of patterns
    observed in a time interval of size $\t$ and in presence of an
    average entropy production $p$ are the same as those of the
    corresponding anti-patterns in presence of the opposite average
    entropy production rate}, \cite[p.476]{Ga02}, or also
    \cite[p.476]{Ga02}, or {\it ... it ``suffices'' to change the sign
    of the entropy production to reverse the arrow of time}, or also
    {\it ... a waterfall will go up, as likely as we see it going
    down, in a world in which for some reason, or by the deed of a
    Daemon, the entropy creation rate has changed sign during a long
    enough time}, \cite[p.288]{Ga00}. We can also say that the motion
    on an attractor is reversible, even in the presence of
    dissipation, once the dissipation is fixed. Again variations of
    this property keep being rediscovered, see for instance \cite{GPB08}. \*

\noindent{}(3) In the case of systems in contact with a single
    thermostat but in a stationary nonequilibrium because of the
    action of external forces the above analysis has also interesting
    consequences. The phase space contraction can be written as
    $\s(x)=\sum_{j>0}\frac{ Q_j+\dot U_j}{k_B T}$, as in
    Eq.(\ref{e1.1}), or by adding to it a time derivative as
    $\overline \s(x)=\s(x)+\b \dot H_0(x)$ which in this case is
    simply $\overline\s(x)=\frac{{\bf E}({\bf X}_0)\cdot\dot{{\bf
    X}}_0}{k_B T}=\frac{\dot W}{k_B T}$. Therefore the fluctuation
    theorem, as pointed out by Bonetto: see \cite[Eq.(9.10.4)]{Ga00},
    yields the following ``{\it work theorem}''

\be \langle{ e^{-\b\, w\, \t}}\rangle_{SRB}=1,
\qquad w{\buildrel def\over=}
\frac1\t\int_0^\t \dot W(S_tx)\, dt\label{e4.1}\ee
in the sense that the logarithm of the l.h.s. divided by $\t$ tends to
$0$ as $\t\to\infty$. More generally the identity up to a time
derivative of $\s$, $\overline\s$, $\sum_{j>0}\frac{Q_j}{k_B T_j}$ and
$\s_0 =\sum_{j>0}(\frac{Q_j}{k_B T_j}-\b Q_j)-\b {\bf E}\cdot\dot{\bf
X}_0$, see Eqs.(\ref{e3.3})-(\ref{e3.6}), implies that, in the same
sense as in Eq.(\ref{e4.1}), the finite time average $P$ of {\it any}
of the latter four quantities, denoted $\widetilde\s$, over a time $\t$ will
satisfy

\be \langle{ e^{-P\t}}\rangle_{SRB}=1,\qquad P{\buildrel def\over=}
  \frac1\t\int_0^\t \widetilde\s(S_tx)\label{e4.2}\ee
which can be called an ``{\it entropy theorem}'': not only remarkable
because it involves quantities that can be measured in experiments,
\cite{Ga08}, but also because here $\b$ can be taken {\it arbitrary},
so that Eq.(\ref{e4.2}) is an infinite number of relations. Actually
if $p=P/\langle{\s_0}\rangle_{SRB}$ the large deviations of $p$ satisfy the
fluctuation theorem symmetry Eq.(\ref{e2.7}). Note however that all
such relations are special cases of the theorem in
\cite{Ga99}. 

\noindent{}(4) A further alternative method to derive the Green-Kubo relations
    is in \cite{CELS93a}. It will be illustrated, for completeness, in
    the simple case of a system interacting with only one thermostat
    and subject to several nonconservative external forces that will
    be proportional to parameters ${\bf E}=(E_1,\ldots,E_q)$. Under
    the chaotic hypothesis the SRB average of the currents $J_m=\int
    \m_{SRB}(dx) j_m(x)$, with $j_m(x){\buildrel
    def\over=}\partial_{E_m}\overline\s(x)$ in presence of
    thermodynamic forces ${\bf E}$, can be computed as the limit
    $J_m=\lim_{t\to\infty} \m_0(j_m(S_t^{{\bf E}}x))$, if $S_t^{\bf E}$
    is the map such that $x\to S^{{\bf E}}_tx$ solves the equations of
    motion in presence of forcing forces with parameters ${\bf E}$,
    and $\m_0$ is the equilibrium distribution Eq.(\ref{e3.1}),
    \cite{CELS93a,CELS93}. Therefore

$$J_m=\lim_{t\to\infty}\m_0(j_m(S_t^{{\bf E}}x))=
\int_0^{+\infty}dt\, \frac{d}{dt} \int \m_0(dx) J_m(S_t^{{\bf E}}x)$$
\be=\int_0^{+\infty}dt\, \frac{d}{dt}\int \frac{\m_0(dx)}{\m_0(dS^{{\bf E}}_tx)}
\m_0(dS^{{\bf E}}_tx) j_m(S^{{\bf E}}_tx)\label{e4.3} \ee
$$=\int_0^{+\infty}dt\, \frac{d}{dt}\int
\frac{\m_0(dS^{{\bf E}}_{-t}x)}{\m_0(dx)}\m_0(dx) j_m(x)$$
but by the comment preceding Eq.(\ref{e3.7}) (considered with
$T_j\equiv T$)

\be \frac{d}{dt} \frac{\m_0(dS^{{\bf E}}_{-t}x)}{\m_0(dx)}=\overline\s(S^{\bf
  E}_{-t}x)\label{e4.4}\ee
so that the chain of equalities in Eq.(\ref{e4.3}) yields

\be J_m=\int_0^\infty dt\,  \int \overline\s(S^{{\bf E}}_{-t}x)
j_m(x)\m_0(dx)\label{e4.5}\ee
And taking into account that $\overline\s(x)\equiv0$, if ${\bf E}={\bf0}$, and
$j_m(x)=\partial_{E_m}\overline\s(x)$

$$L_{pm}=\partial_{E_p} J_m|_{{\bf E}={\bf0}}= \int_0^\infty
dt\,\big( \int \partial_{E_p}\overline\s( S^{\bf
E}_{-t}x)\,\partial_{E_m}\overline\s(x)\, \m_0(dx)\big)
\big|_{{\bf E}={\bf0}}$$
\be=\frac12 \int_{-\infty}^\infty dt \big(\int
\partial_{E_p}\overline\s( S^{\bf
  E}_{-t}x)\,\partial_{E_m}\overline\s(x)\,\m_0(dx)\big)
\big|_{\bf E}= {\bf 0}=L_{mp}\label{e4.6}\ee
by time reversal invariance of the equilibrium distribution $\m_0$,
which is the Green-Kubo formula.
%\bibliography{0Bib}

\begin{thebibliography}{10}

\bibitem{No84}
S.~Nos\'e.
\newblock A unified formulation of the constant temperature molecular dynamics
  methods.
\newblock {\em Journal of Chemical Physics}, 81:511--519, 1984.

\bibitem{Ho85}
W.~Hoover.
\newblock Canonical equilibrium phase-space distributions.
\newblock {\em Physical Review A}, 31:1695--1697, 1985.

\bibitem{EM90}
D.~J. Evans and G.~P. Morriss.
\newblock {\em Statistical Mechanics of Non{\-}equilibrium Fluids}.
\newblock Academic Press, New-York, 1990.

\bibitem{Ga08}
G.~Gallavotti.
\newblock Heat and fluctuations from order to chaos.
\newblock {\em European Physics Journal B, EPJB}, 61:1--24, 2008.

\bibitem{Ga07b}
G.~Gallavotti.
\newblock Fluctuation relation, fluctuation theorem, thermostats and entropy
  creation in non equilibrium statistical physics.
\newblock {\em C.R. Physique}, 8:486--494, 2007,
  (doi:10.1016/j.crhy.2007.04.011).

\bibitem{BGGZ05}
F.~Bonetto, G.~Gallavotti, A.~Giuliani, and F.~Zamponi.
\newblock Chaotic {H}ypothesis, {F}luctuation {T}heorem and {S}ingularities.
\newblock {\em Journal of Statistical Physics}, 123:39--54, 2006.

\bibitem{Be64}
R.~Becker.
\newblock {\em Electromagnetic fields and interactions}.
\newblock Blaisdell, New-York, 1964.

\bibitem{CELS93}
N.~I. Chernov, G.~L. Eyink, J.~L. Lebowitz, and {Ya.}~G. Sinai.
\newblock Steady state electric conductivity in the periodic {L}orentz gas.
\newblock {\em Communications in Mathematical Physics}, 154:569--601, 1993.

\bibitem{GG07}
P.~Garrido and G.~Gallavotti.
\newblock Boundary dissipation in a driven hard disk system.
\newblock {\em Journal of Statistical Physics}, 126:1201--1207, 2007.

\bibitem{ECM93}
D.~J. Evans, E.~G.~D. Cohen, and G.~P. Morriss.
\newblock Probability of second law violations in shearing steady flows.
\newblock {\em Physical Review Letters}, 71:2401--2404, 1993.

\bibitem{GC95}
G.~Gallavotti and E.~G.~D. Cohen.
\newblock Dynamical ensembles in nonequilibrium statistical mechanics.
\newblock {\em Physical Review Letters}, 74:2694--2697, 1995.

\bibitem{Ga00}
G.~Gallavotti.
\newblock {\em Statistical Mechanics. A short treatise}.
\newblock Springer Verlag, Berlin, 2000.

\bibitem{Ga96}
G.~Gallavotti.
\newblock Chaotic hypothesis: {O}nsager reciprocity and
  fluctuation--dissi\-pation theorem.
\newblock {\em Journal of Statistical Physics}, 84:899--926, 1996.

\bibitem{Ja99}
C.~Jarzynski.
\newblock Hamiltonian derivation of a detailed fluctuation theorem.
\newblock {\em Journal of Statistical Physics}, 98:77--102, 1999.

\bibitem{BK81a}
G.~N. Bochkov and Yu.~E. Kuzovlev.
\newblock Nonlinear fluctuation-dissipation relations and stochastic models in
  nonequilibrium thermodynamics: I. generalized fluctuation-dissipation
  theorem.
\newblock {\em Physica A}, 106:443--479, 1981.

\bibitem{Ga99}
G.~Gallavotti.
\newblock New methods in nonequilibrium gases and fluids.
\newblock {\em Open Systems and Information Dynamics}, 6:101--136, 1999
  (preprint chao-dyn/9610018).

\bibitem{Ru78b}
D.~Ruelle.
\newblock What are the measures describing turbulence.
\newblock {\em Progress in Theoretical Physics Supplement}, 64:339--345, 1978.

\bibitem{Ru98}
D.~Ruelle.
\newblock Natural nonequilibrium states in quantum statistical mechanics.
\newblock {\em Journal of Statistical Physics}, 98:57--75, 1998.

\bibitem{Ru73}
D.~Ruelle.
\newblock {\em Ergodic theory}, volume Suppl X of {\em The Boltzmann equation,
  ed. E.G.D Cohen, W. Thirring, Acta Physica Austriaca}.
\newblock Springer, New York, 1973.

\bibitem{Bo866}
L.~Boltzmann.
\newblock {\em {\"U}ber die mechanische {B}edeutung des zweiten {H}auptsatzes
  der {W\"a}rme\-theorie}, volume 1, p.9 of {\em {W}is\-sen\-schaft\-li\-che
  {A}bhandlungen, ed. {F}. {H}asen{\"o}hrl}.
\newblock Chelsea, New York, 1968.

\bibitem{Cl871}
R.~Clausius.
\newblock Ueber die zur{\"u}ckf{\"u}hrung des zweites hauptsatzes der
  mechanischen w{\"a}rmetheorie aud allgemeine mechanische prinzipien.
\newblock {\em Annalen der Physik}, 142:433--461, 1871.

\bibitem{He884a}
H.~Helmholtz.
\newblock {\em Prinzipien der Statistik monocyklischer Systeme}, volume III of
  {\em {W}is\-sen\-schaft\-li\-che {A}bhandlungen}.
\newblock Barth, Leipzig, 1895.

\bibitem{MR007}
C.~Mejia-Monasterio and L.Rondoni.
\newblock On the fluctuation relation for {N}ose-{H}oover locally thermostated
  systems.
\newblock {\em arXiv: cond-mat}, 0710.3673:1--27, 2007.

\bibitem{Si68a}
{Ya.}~G. Sinai.
\newblock Markov partitions and {$C$}-diffeomorphisms.
\newblock {\em Functional Analysis and Applications}, 2(1):64--89, 1968.

\bibitem{Bo70a}
R.~Bowen.
\newblock Markov partitions for axiom {A} diffeomorphisms.
\newblock {\em American Journal of Mathematics}, 92:725--747, 1970.

\bibitem{BR75}
R.~Bowen and D.~Ruelle.
\newblock The ergodic theory of axiom {A} flows.
\newblock {\em Inventiones Mathematicae}, 29:181--205, 1975.

\bibitem{Ge98}
G.~Gentile.
\newblock A large deviation theorem for {A}nosov flows.
\newblock {\em Forum Mathematicum}, 10:89--118, 1998.

\bibitem{Si77}
{Ya.}~G. Sinai.
\newblock {\em Lectures in ergodic theory}.
\newblock Lecture notes in Mathematics. Princeton University Press, Princeton,
  1977.

\bibitem{GBG04}
G.~Gallavotti, F.~Bonetto, and G.~Gentile.
\newblock {\em Aspects of the ergodic, qualitative and statistical theory of
  motion}.
\newblock Springer Verlag, Berlin, 2004.

\bibitem{GC95b}
G.~Gallavotti and E.G.D. Cohen.
\newblock Dynamical ensembles in stationary states.
\newblock {\em Journal of Statistical Physics}, 80:931--970, 1995.

\bibitem{Ga95b}
G.~Gallavotti.
\newblock Reversible {A}nosov diffeomorphisms and large deviations.
\newblock {\em Mathematical Physics Electronic Journal (MPEJ)}, 1:1--12, 1995.

\bibitem{CG99}
E.~G.~D. Cohen and G.~Gallavotti.
\newblock Note on two theorems in nonequilibrium statistical mechanics.
\newblock {\em Journal of Statistical Physics}, 96:1343--1349, 1999.

\bibitem{Ga97}
G.~Gallavotti.
\newblock Fluctuation patterns and conditional reversibility in nonequilibrium
  systems.
\newblock {\em Annales de l' Institut H. Poincar\'e}, 70:429--443, 1999 and
  chao-dyn/9703007.

\bibitem{Ga02}
G.~Gallavotti.
\newblock {\em Foundations of Fluid Dynamics}.
\newblock (second printing) Sprin\-ger Verlag, Berlin, 2005.

\bibitem{GR97}
G.~Gallavotti and D.~Ruelle.
\newblock {SRB} states and non\-equi\-li\-brium statistical mechanics close to
  equi\-li\-brium.
\newblock {\em Com\-mu\-ni\-ca\-tions in Mathematical Physics}, 190:279--285,
  1997.

\bibitem{Ru96}
D.~Ruelle.
\newblock Positivity of entropy production in nonequilibrium statistical
  mechanics.
\newblock {\em Journal of Statistical Physics}, 85:1--25, 1996.

\bibitem{Ga96a}
G.~Gallavotti.
\newblock Extension of {O}nsager's reciprocity to large fields and the chaotic
  hypothesis.
\newblock {\em Physical Review Letters}, 77:4334--4337, 1996.

\bibitem{GPB08}
A.~Gomez-Marin, J.M.R. Parondo, and C.~Van den Broeck.
\newblock The footprints of irreversibility.
\newblock {\em European Physics Letters}, 82:5002+4, 2008.

\bibitem{CELS93a}
N.~I. Chernov, G.~L. Eyink, J.~L. Lebowitz, and {Ya.}~G. Sinai.
\newblock Derivation of {O}hm's law in a deterministic mechanical model.
\newblock {\em Physical Review Letters}, 70:2209--2212, 1993.

\end{thebibliography}
 \bibliographystyle{unsrt}

\end{document}